\documentclass[12pt,draftcls,onecolumn]{IEEEtran}
\usepackage[utf8]{inputenc}
\usepackage{easyReview} 
\usepackage{soulutf8} 
\usepackage{amsmath,mathtools}
\usepackage{amsthm}
\usepackage{amsfonts}
\usepackage{amssymb}
\usepackage{amsmath}
\usepackage{flexisym}
\usepackage{breqn}
\usepackage{pgfkeys}
\usepackage{array}
\DeclareUnicodeCharacter{00A0}{ }

\usepackage[justification=centering]{caption}
\usepackage{enumerate} 
\usepackage{cite}
\usepackage{placeins}
\newcommand{\vast}{\bBigg@{4}}
\newcommand{\Vast}{\bBigg@{5}}

\usepackage{tikz}
\theoremstyle{definition}

\DeclarePairedDelimiter\curlybracket{\lbrace}{\rbrace}
\allowdisplaybreaks
\newcolumntype{C}[1]{>{\centering\let\newline\\\arraybackslash\hspace{0pt}}m{#1}}
\begin{document}
	\bstctlcite{IEEEexample:BSTcontrol}
	\title{A New Expression for the Product of Two $\kappa - \mu $ Shadowed Random Variables and its Application to Wireless Communication}
	\author{Shashank Shekhar, Sheetal Kalyani \\
		\hspace{-0.5 cm}Department of Electrical Engineering,\\
		\hspace{-0.8 cm} Indian Institute of Technology, Madras, \\
		\hspace{-1cm} Chennai, India 600036.\\
		\hspace{-1cm} \{ee17d022@smail,skalyani@ee\}.iitm.ac.in\\
	}
	\maketitle
	\begin{abstract}
		In this work, the product of two independent and non-identically distributed (i.n.i.d) $ \kappa - \mu $ shadowed random variables is studied. We derive the series expression for the probability density function (PDF), cumulative distribution function (CDF), and moment generating function (MGF) of the product of two (i.n.i.d) $ \kappa - \mu $ shadowed random variables. The derived formulation in this work is quite general as they incorporate most of the typically used fading channels. As an application example, outage probability (OP) has been derived for cascaded wireless systems and relay-assisted communications with a variable gain relay. Extensive Monte-Carlo simulations have also been carried out.
	\end{abstract}
	\section{Introduction}
	A wireless channel is governed mainly by two physical phenomena shadowing (which results in long-term signal variation) and multipath (which results in short-term fading). Shadowing is typically modeled using lognormal distribution \cite{alouini2002dual} or sometimes using gamma distribution \cite{abdi1999utility}. In contrast to shadowing, multipath effect is characterized by a broad range of distribution such as Rayleigh, Rician, Nakagami-$ m $, Hoyt and more general distribution like $ \kappa-\mu $, $ \eta-\mu $ \cite{Yacoub2007:KappaMu} and $ \alpha-\mu $ \cite{yacoub2007alpha}. The $  \kappa - \mu $ shadowed distribution introduced in \cite{paris2013:statistical} provides a natural generalization of the $ \kappa - \mu $ fading where the line-of-sight (LOS) component of received signal is random in nature, \textit{i.e.}, subject to shadowing. This type of fading model is known as the LOS shadow fading model in the literature. In \cite{lopez2017kappa} the authors showed that the   $ \kappa - \mu $ shadowed distribution with an integer value of $ \mu $ and $ m $ can be represented as a mixture of Gamma distribution.
	\par In a variety of wireless communication applications, such as relay-based communication systems \cite{talha2011channel},  and, intelligent reflecting surfaces (IRS) assisted communication system \cite{lis_renzo_2020,wu2019towards,Zhang5g}, the transmitted signal from the source reaches the destination after experiencing a couple of fading environments. To analyze such a communication system's performance, one needs to know the statistics of the product of corresponding fading distributions. Hence, the statistical characterization of the product of two random variables (RVs) has crucial importance in wireless communication. 
	For example, in bistatic scatter radio communication, the indirect channel between carrier emitter and software-defined radio (SDR) reader through a radio frequency (RF)  tag is modeled as the product of two Rayleigh and Rician fading channels in \cite{fasarakis2015coherent} and \cite{alevizos2017noncoherent}, respectively. The work in \cite{fasarakis2015coherent} and \cite{alevizos2017noncoherent} were focused on point to point communication, whereas a multiscatter scenario is considered in \cite{alevizos2018multistatic} where multiple carrier emitters are present, and each channel is modeled as Nakagami$-m$ fading. Hence, the channel between carrier emitter to SDR  reader is a product of two Nakagami$-m$ RVs. In \cite{salo2006distribution}, authors presented a general result for the product statistics of Rayleigh fading. A generic cascaded channel has been considered in \cite{karagiannidis2007n,yilmaz2009product} with Nakagami$-m$ and generalized Nakagami$-m$ fading, respectively. Authors in \cite{Bhargav2018:ProductKappaMu} studied the product statistics of two independent
	and non-identically distributed $ \kappa-\mu $ RVs. In \cite{Silva:ProductAlphaKappaEtaMu} the statistical characterization of $ \alpha-\mu $ and $ \eta-\mu $ RV is done along with $ \kappa-\mu $ RV.  
	Recently, the authors in \cite{Charishma2021OutageIRS} considered an IRS-assisted communication system where each link undergoes $ \kappa-\mu $ fading, and hence the link between source and destination via IRS is the product of two $ \kappa-\mu $ RVs.   
	\par In this work, we are interested in $ \kappa - \mu $ shadowed fading since it unites various popular fading models such as one-sided Gaussian, Rician, Rayleigh, $ \kappa - \mu $, Nakagami-$m$ and Rician shadowed. Apart from its generalized nature, $ \kappa - \mu $ shadowed distribution has good analytical tractability and it found a lot of traction in wireless community in recent literature \cite{kumar2015approximate,kumar2017outage,chandrasekaran2015performance,bhatnagar2014sum,cotton2015second,srinivasan2018secrecy,subhash2019asymptotic}. This motivates us to look at the product statistics of $ \kappa - \mu $ shadowed fading. We provided the statistical characterization of the product of two independent non-identically distributed (i.n.i.d.)  $ \kappa - \mu $ shadowed RVs. The closed-form exact expression for PDF and CDF of the considered RV is derived using Mellin transformation.     
	The contribution and utility of this work are summarized as follows: 
	\begin{itemize}
		\item Series expressions for PDF, CDF, and MGF of the product of two, i.n.i.d. $ \kappa-\mu $ shadowed RV are derived using the direct application of Mellin transform \footnote{Very recently, some statistics of the product of two $ \kappa-\mu $ shadowed random variable has been derived in \cite{bilim2022cascaded}. However, the method utilized there is totally different from the one we used in this work. Hence, the resulting expression are also different and original. Derived series expression involve simple hypergeometric function hence can be easily computed using popular software as Mathematica.}.
		\item We presented the performance metrics for a cascaded wireless system and outage probability for a relay-assisted wireless communication system as an application example for the presented fading distribution.     
	\end{itemize}

	\section{Proposed Statistical Characterization Using Inverse Mellin Transform}\label{Sec:ProductKMS_Mellin}
		We considered two independent non-identically distributed (i.n.i.d) $ \kappa-\mu $ shadowed RVs, say $ X_{i} $ with mean $ \bar{\gamma}_{i} $ and non-negative real shaping parameters $ \kappa_{i},\mu_{i}, m_{i} $ for $ i = 1,2 $. Each $ X_{i} $ follows the distribution given by\cite[eq. $(4)$]{paris2013:statistical},
	\begin{equation}
		\label{Eq:KappaMupdf}
		\begin{aligned}
			f_{X_{i}}(x_{i}) &=\frac{\mu_{i}^{\mu_{i}} m_{i}^{m_{i}}(1+\kappa_{i})^{\mu_{i}}}{\Gamma(\mu_{i}) \bar{\gamma}_{i}(\mu_{i} \kappa_{i}+m_{i})^{m_{i}}}\left(\frac{x_{i}}{\bar{\gamma}_{i}}\right)^{\mu_{i}-1}  e^{-\frac{\mu_{i}(1+\kappa_{i}) x_{i}}{\bar{\gamma}_{i}}}  { }_{1} F_{1}\left(m_{i}; \mu_{i} ; \frac{\mu_{i}^{2} \kappa_{i}(1+\kappa_{i})}{\mu_{i} \kappa_{i} + m_{i}} \frac{x_{i}}{\bar{\gamma}_{i}}\right),
		\end{aligned}
	\end{equation}
	where $ i = 1,2 $. $ \kappa_{i} $ is the ratio of power contribution from dominant path to scattered waves, $ \mu_{i} $ is the real extension to number of multipath clusters and $ m_{i} $ is the shaping parameter for LOS shadowing component, $ {}_{1} F_{1}\left(\cdot;\cdot;\cdot\right) $ is confluent hypergeometric function\cite{srivastava1985:multiple}. 
	\par Our objective is to statistically characterize the RV  $ Y = X_{1} X_{2} $.         
	We will now use the technique of Mellin transform to derive the PDF of $ Y $. First, we re-write the PDF of $ X_{i} $ as follows
	\begin{equation}
		\begin{aligned}
			f_{X_{i}}(x_{i})= \theta_{i} x_{i}^{\mu_{i}-1} g_{i}\left(x_{i}\right),
		\end{aligned}
	\end{equation} 
	where $ \theta_{i} = \frac{ a_{i}^{\mu_{i}} b_{i}}{\Gamma(\mu_{i})} , g_{i}\left(x_{i}\right) =  e^{- a_{i}x_{i}} { }_{1} F_{1}\left(m_{i}, \mu_{i} ; a_{i}c_{i} x_{i}\right), a_{i} = \frac{\mu_{i}\left(1 + \kappa_{i}\right)}{\bar{\gamma}_{i}}, b_{i} = \frac{m_{i}^{m_{i}}}{\left(\mu_{i} \kappa_{i}+m_{i}\right)^{m_{i}}} $ and $  c_{i} = \frac{ \mu_{i} \kappa_{i}}{\left(\mu_{i} \kappa_{i} + m_{i}\right)} $. Then, the Mellin transform of $ f_{X_{i}}(x_{i}) $ is 
	\begin{equation}
		\begin{aligned}
			\mathcal{M}\left[f_{X_{i}}(x_{i});s\right] = \theta_{i}\mathcal{M}\left[g_{i}(x_{i});s+\mu_{i}-1\right]
		\end{aligned}
	\end{equation}
	Now, we need to find the Mellin transform of $ g_{i}\left(x_{i}\right) $ which is 
	\begin{equation}
		\begin{aligned}
			\mathcal{M}\left[g_{i}(x_{i});s\right] &= \int_{0}^{\infty} x_{i}^{s-1} g_{i}\left(x_{i}\right) d x_{i} \\
			&= \int_{0}^{\infty} x_{i}^{s-1} e^{- a_{i}x_{i}} { }_{1} F_{1}\left(m_{i}, \mu_{i} ; a_{i}c_{i} x_{i}\right) dx_{i}
		\end{aligned}
	\end{equation}
	Using the identity \cite[eq. $(7.621.4)$]{Grad2007}, we have
	\begin{equation}
		\begin{aligned}
			\mathcal{M}\left[g_{i}(x_{i});s\right] &= \Gamma\left(s\right) a_{i}^{-s} {}_{2}F_{1}\left(m_{i},s;\mu_{i};c_{i}\right)
		\end{aligned}
	\end{equation}
	Finally,
	\begin{equation}
		\begin{aligned}
			\mathcal{M}\left[f_{X_{i}}(x_{i});s\right] &= \theta_{i}\frac{\Gamma\left(s+\mu_{i}-1\right)}{ a_{i}^{s+\mu_{i}-1}} {}_{2}F_{1}\left(m_{i},s+\mu_{i}-1;\mu_{i};c_{i}\right) \\
			&= \frac{b_{i}\Gamma\left(s+\mu_{i}-1\right)}{\Gamma(\mu_{i}) a_{i}^{s-1}} {}_{2}F_{1}\left(m_{i},s+\mu_{i}-1;\mu_{i};c_{i}\right)
		\end{aligned}
	\end{equation}
	It is a well-known fact that the Mellin convolution of individual PDFs gives the PDF of the product of two independent RVs, and the Mellin transform of the said PDF is the product of the Mellin transform of corresponding PDFs \cite{springer1979:RValgebra}. Hence, the Mellin transform of $ Y $ is 
	\begin{equation}\label{Eq:MellinTransY}
		\begin{aligned}
			\mathcal{M}\left[f_{Y}\left(y\right);s\right] &= \prod_{i=1}^{2}\mathcal{M}\left[f_{X_{i}}(x_{i});s\right] \\
			&= \frac{b_{1}b_{2}\Gamma\left(s+\mu_{1}-1\right)\Gamma\left(s+\mu_{2}-1\right)}{\Gamma(\mu_{1})\Gamma(\mu_{2}) \left(a_{1}a_{2}\right)^{s-1}}
			{}_{2}F_{1}\left(m_{1},s+\mu_{1}-1;\mu_{1};c_{1}\right) \\&
			\times {}_{2}F_{1}\left(m_{2},s+\mu_{2}-1;\mu_{2};c_{2}\right)
		\end{aligned}
	\end{equation}
	Now, $ f_{Y}\left(y\right) $ is obtained using the inverse Mellin transform, \textit{i.e.},
	\begin{equation}
		\begin{aligned}
			f_{Y}\left(y\right) &= \frac{1}{2\pi i} \int_{c-i\infty}^{c + i\infty} \mathcal{M}\left[f_{Y}\left(y\right);s\right] y^{-s} ds \\ 
			&= \frac{1}{2\pi i} \int_{c-i\infty}^{c + i\infty} \Bigg(
			\frac{b_{1}b_{2}\Gamma\left(s+\mu_{1}-1\right)\Gamma\left(s+\mu_{2}-1\right)}{\Gamma(\mu_{1})\Gamma(\mu_{2}) \left(a_{1}a_{2}\right)^{s-1}} 
			{}_{2}F_{1}\left(m_{1},s+\mu_{1}-1;\mu_{1};c_{1}\right) \Bigg.
			\\&\hspace{30mm} \Bigg. 
			\times {}_{2}F_{1}\left(m_{2},s+\mu_{2}-1;\mu_{2};c_{2}\right)y^{-s}\Bigg) ds 
		\end{aligned}
	\end{equation}
	By definition, $ c_{1},c_{2} < 1 $ and $ \mu_{1},\mu_{2} > 0 $ so both hypergeometric function, present in the integrand of above integral, will be analytic $ \forall s $. Hence, value of the integral will be decided by the position of poles of Gamma function. Based on the value of $ \mu_{1},\mu_{2} $, there are two cases. 
	\subsubsection{When $ \mu_{2} - \mu_{1} \notin \mathbb{Z} $}
	In this case, the poles of $ \Gamma(s + \mu_{1} -1) $ and $ \Gamma(s + \mu_{2} -1) $ are distinct. Hence, by the virtue of residue theorem and Jordan's Lemma\cite{springer1979:RValgebra}, we have
	\begin{equation}
		\begin{aligned}
			f_{Y}\left(y\right) &= \frac{b_{1}b_{2}}{\Gamma(\mu_{1})\Gamma(\mu_{2})} \sum_{n=0}^{\infty} \left[ R_{1,n}+ R_{2,n}\right],
		\end{aligned}
	\end{equation}
	where 
	\begin{equation}
		\begin{aligned}
			R_{1,n} &= \lim_{s \rightarrow -n-\mu_{1}+1}
			\frac{\left(s+n+\mu_{1}-1\right)\Gamma\left(s+\mu_{1}-1\right)\Gamma\left(s+\mu_{2}-1\right)}{ \left(a_{1}a_{2}\right)^{s-1}}  {}_{2}F_{1}\left(m_{1},s+\mu_{1}-1;\mu_{1};c_{1}\right) 
			\\& \times {}_{2}F_{1}\left(m_{2},s+\mu_{2}-1;\mu_{2};c_{2}\right) y^{-s} \\ 
			&= 	\frac{\left(a_{1}a_{2}\right)^{n+\mu_{1}}\Gamma\left(\mu_{2}-\mu_{1}-n\right)}{ (-1)^{n}n!} {}_{2}F_{1}\left(m_{1},-n;\mu_{1};c_{1}\right)  {}_{2}F_{1}\left(m_{2},\mu_{2}-\mu_{1}-n;\mu_{2};c_{2}\right) y^{n+\mu_{1}-1}
		\end{aligned}
	\end{equation} 
	and, similarly
	\begin{equation}
		\begin{aligned}
			R_{2,n}	&= 	\frac{\left(a_{1}a_{2}\right)^{n+\mu_{2}}\Gamma\left(\mu_{1}-\mu_{2}-n\right)}{ (-1)^{n}n!}
			{}_{2}F_{1}\left(m_{1},\mu_{1} - \mu_{2}-n;\mu_{1};c_{1}\right)  {}_{2}F_{1}\left(m_{2},-n;\mu_{2};c_{2}\right) y^{n+\mu_{2}-1}
		\end{aligned}
	\end{equation} 
	Hence, PDF of $ Y $ is 
	\begin{equation}\label{Eq:ExactPDFMellin1}
		\begin{aligned}
			f_{Y}\left(y\right) &= \frac{b_{1}b_{2}}{\Gamma(\mu_{1})\Gamma(\mu_{2})} \sum_{n=0}^{\infty}\left[A_{n}y^{n+\mu_{1}-1} + B_{n} y^{n+\mu_{2}-1}\right],
		\end{aligned}
	\end{equation}
	where $ A_{n} = \frac{\left(a_{1}a_{2}\right)^{n+\mu_{1}}\Gamma\left(\mu_{2}-\mu_{1}-n\right)}{ (-1)^{n}n!} {}_{2}F_{1}\left(m_{1},-n;\mu_{1};c_{1}\right) \allowbreak {}_{2}F_{1}\left(m_{2},\mu_{2}-\mu_{1}-n;\mu_{2};c_{2}\right)  $ and $ B_{n} = \frac{\left(a_{1}a_{2}\right)^{n+\mu_{2}}\Gamma\left(\mu_{1}-\mu_{2}-n\right)}{ (-1)^{n}n!}
	{}_{2}F_{1}\left(m_{1},\mu_{1} - \mu_{2}-n;\mu_{1};c_{1}\right)  {}_{2}F_{1}\left(m_{2},-n;\mu_{2};c_{2}\right) $. Note that the $ A_{n} $ and $ B_{n} $ only depends on the parameters of both $ \kappa-\mu $ shadowed RV \textit{i.e.}, independent of $ y $. Thus, the PDF of $ Y $ is simply a power series of $ y $.   
	\subsubsection{When $ \mu_{2} - \mu_{1} \in \mathbb{Z} $} Without loss of generality, let $ \mu_{2} > \mu_{1} $ and $ \mu_{2} - \mu_{1} = N $ then the poles of $ \Gamma(s + \mu_{1} -1) $ and $ \Gamma(s + \mu_{2} -1) $ coincides for $ n \ge N $. So, there are $ N $ poles of order one and the remaining poles are of order two. Again, using residue theorem, we have  
	\begin{equation}
		\begin{aligned}
			f_{Y}\left(y\right) &= \frac{b_{1}b_{2}}{\Gamma(\mu_{1})\Gamma(\mu_{2})} \left( \sum_{n=0}^{N-1} S_{1,n} +  \sum_{n=N}^{\infty} S_{2,n} \right).
		\end{aligned}
	\end{equation}
	The first $ N $ poles are due to $ \Gamma(s + \mu_{1} -1) $ so we have 
	\begin{equation}
		\begin{aligned}
			S_{1,n} &= A_{n} y^{n+\mu_{1}-1}, \quad  n = 0,1,\dots,N-1.
		\end{aligned}
	\end{equation} 
	and 
	\begin{equation}
		\begin{aligned}
			S_{2,n} &= \frac{\left(-1\right)^{N} \left(a_{1}a_{2}\right)^{n+\mu_{1}} y^{n+\mu_{1}-1} }{\left(n - N\right)! n!} \Bigg\{ {}_{2}F_{1}^{\left(0,1,0,0\right)}\left(m_{1},-n;\mu_{1};c_{1}\right)  {}_{2}F_{1}\left(m_{2},-n+N;\mu_{2};c_{2}\right) \\
			&\hspace{-5mm}+ {}_{2}F_{1}\left(m_{1},-n;\mu_{1};c_{1}\right)  {}_{2}F_{1}^{\left(0,1,0,0\right)}\left(m_{2},-n+N;\mu_{2};c_{2}\right) \\
			&\hspace{-5mm}+ \left[\psi\left(n+1\right) + \psi\left(n-N+1\right) - \ln\left(y\right) - \ln\left(a_{1}a_{2}\right)\right] 
			{}_{2}F_{1}\left(m_{1},-n;\mu_{1};c_{1}\right)  {}_{2}F_{1}\left(m_{2},-n+N;\mu_{2};c_{2}\right) \Bigg\}.
		\end{aligned}
	\end{equation}
	details for the calculation of $ S_{2,n} $ for $ n \ge N $ is given in Appendix \ref{App:ProductKMS_S2}. Finally, the PDF of $ Y $ for the case when $ \mu_{2} - \mu_{1} \in \mathbb{Z} $ is 
	\begin{equation}
		\begin{aligned}\label{Eq:ExactPDFMellin2}
			f_{Y}\left(y\right) &= 	\frac{b_{1}b_{2}}{\Gamma(\mu_{1})\Gamma(\mu_{2})}  \left(\sum_{n=0}^{N-1}
			A_{n} y^{n+\mu_{1}-1} + \sum_{n=N}^{\infty} \left[ C_{n} - D_{n} \ln\left(y\right)\right] y^{n+\mu_{1}-1}\right),
		\end{aligned}
	\end{equation} 
	where 
	\begin{equation}\label{Eq:DefK3n}
		\begin{aligned}
			C_{n}	&= \frac{\left(-1\right)^{N} \left(a_{1}a_{2}\right)^{n+\mu_{1}}  }{\left(n - N\right)! n!} \Bigg\{ {}_{2}F_{1}^{\left(0,1,0,0\right)}\left(m_{1},-n;\mu_{1};c_{1}\right)   {}_{2}F_{1}\left(m_{2},-n+N;\mu_{2};c_{2}\right)\Bigg. \\
			&+ {}_{2}F_{1}\left(m_{1},-n;\mu_{1};c_{1}\right)  {}_{2}F_{1}^{\left(0,1,0,0\right)}\left(m_{2},-n+N;\mu_{2};c_{2}\right) \\
			&+ \left[\psi\left(n+1\right) + \psi\left(n-N+1\right) - \ln\left(a_{1}a_{2}\right)\right]
			{}_{2}F_{1}\left(m_{1},-n;\mu_{1};c_{1}\right) 
			{}_{2}F_{1}\left(m_{2},-n+N;\mu_{2};c_{2}\right) \Bigg\}.
		\end{aligned}
	\end{equation}
	and,
	\begin{equation}
		\begin{aligned}
			D_{n}	&= \frac{\left(-1\right)^{N} \left(a_{1}a_{2}\right)^{n+\mu_{1}}  }{\left(n - N\right)! n!} {}_{2}F_{1}\left(m_{1},-n;\mu_{1};c_{1}\right)
			{}_{2}F_{1}\left(m_{2},-n+N;\mu_{2};c_{2}\right).
		\end{aligned}
	\end{equation}
	In (\ref{Eq:DefK3n}), $\psi(.)$ is the digamma function \cite{Polygamma} and ${}_2 F_1^{(0,1,0,0)}(a,b,c,z)$ is the derivative of the confluent hypergeometric function with respect to the parameter $b$ \cite{diff2F1wrtb}
	\begin{equation}
		\begin{aligned}
		{}_2 F_1^{(0,1,0,0)}(a,b,c,z)	&=  \frac{z a}{c}F^{2,1,2}_{2,0,1}\left(\begin{array}{c|}
			a+1,b+1 ; 1 ; 1,b \\
			2,c+c ; - ; b+1 
		\end{array} \ z,z\right),
		\end{aligned}
	\end{equation}
	where $ F^{p,q,k}_{l,m,n}\left(\begin{array}{c|}
		\left(\mathbf{a}_{p}\right) ; \left(\mathbf{b}_{q}\right) ; \left(\mathbf{c}_{k}\right) \\
		\left(\boldsymbol{\alpha}_{l}\right) ; \left(\boldsymbol{\beta}_{m}\right) ; \left(\boldsymbol{\gamma}_{n}\right)  
	\end{array} \ x,y\right) $ is the Kamp\'e de F\'eriet's Series.
	\subsection{Cumulative Distribution Function}
	Similar to PDF, CDF expression also depends the values of $ \mu_{1} $ and $ \mu_{2} $. Using PDF in (\ref{Eq:ExactPDFMellin1}) the CDF for the case when $ \mu_{2} -\mu_{1} \notin \mathbb{Z} $ is given as 
	\begin{equation} \label{Eq:ExactCDFMellin1}
		\begin{aligned}
			F_{Y}\left(y\right) &= \frac{b_{1}b_{2}}{\Gamma(\mu_{1})\Gamma(\mu_{2})} \sum_{n=0}^{\infty}\left[\frac{A_{n}}{n+\mu_{1}}y^{n+\mu_{1}} + \frac{B_{n}}{n+\mu_{2}}y^{n+\mu_{2}}\right]
		\end{aligned}
	\end{equation}
	and the CDF for the other case, using PDF in (\ref{Eq:ExactPDFMellin2}) is as follows
	\begin{equation}
		\begin{aligned}\label{Eq:ExactCDFMellin2}
			F_{Y}\left(y\right) &= 	\frac{b_{1}b_{2}}{\Gamma(\mu_{1})\Gamma(\mu_{2})}  \left( \sum_{n=0}^{N-1} \frac{A_{n}}{n+\mu_{1}}y^{n+\mu_{1}} +  \sum_{n=N}^{\infty} \frac{\left[C_{n} - D_{n}\ln\left(y\right)\right]}{n+\mu_{1}}y^{n+\mu_{1}}  + \sum_{n=N}^{\infty} \frac{D_{n}}{\left(n + \mu_{1}\right)^{2}}y^{n+\mu_{1}} \right)
		\end{aligned}
	\end{equation}
	Now, we derive the expression for MGF using the PDF in (\ref{Eq:ExactPDFMellin1}) and (\ref{Eq:ExactPDFMellin2}).
	\subsection{Moment Generating Function}
	The MGF of $ Y $ using the PDF in (\ref{Eq:ExactPDFMellin1}) is given as
	\begin{equation}
		\begin{aligned}
			M_{Y}\left(s\right)	&= \mathcal{L}\left[f_{Y}\left(y\right);-s\right] \\
			&= \frac{b_{1}b_{2}}{\Gamma(\mu_{1})\Gamma(\mu_{2})} \sum_{n=0}^{\infty}\left[ \frac{A_{n}\Gamma\left(n+\mu_{1}\right)}{\left(-s\right)^{n+\mu_{1}}} + \frac{B_{n}\Gamma\left(n+\mu_{2}\right)}{\left(-s\right)^{n+\mu_{2}}}
			\right]
		\end{aligned}
	\end{equation}
	and the MGF for the case when $ \mu_{2} - \mu_{1} \in \mathbb{Z} $, using PDF in (\ref{Eq:ExactPDFMellin2}) is as follows
	\begin{equation}
		\begin{aligned}
			M_{Y}\left(s\right)	&=  \frac{b_{1}b_{2}}{\Gamma(\mu_{1})\Gamma(\mu_{2})}  \left( \sum_{n=0}^{N-1} \frac{A_{n}\Gamma\left(n+\mu_{1}\right) }{\left(-s\right)^{n+\mu_{1}}}  +  \sum_{n=N}^{\infty} \frac{\left[C_{n} + D_{n}\left(\ln\left(-s\right) - \psi\left(n+\mu_{1}\right)\right) \right]\Gamma\left(n+\mu_{1}\right) }{\left(-s\right)^{n+\mu_{1}}}\right)
		\end{aligned}
	\end{equation}
	\subsection{Moments}
	The $ n $-th  order moment of RV $ Y $ is given by $ \mathbb{E}\left[Y^{n}\right] = \mathbb{E}\left[X_{1}^{n}\right]\mathbb{E}\left[X_{2}^{n}\right]$, since $ X_{1} $ and $ X_{2} $ are independent RVs. Also, note that the $ \mathbb{E}\left[Y^{n}\right] = \mathcal{M}\left[f_{Y}\left(y\right);s\right]\vert_{s = n+1} $. Hence, from (\ref{Eq:MellinTransY}) we have
	\begin{equation}\label{Eq:Moments}
		\begin{aligned}
			\mathbb{E}\left[Y^{n}\right] &= 
			\frac{b_{1}b_{2}\Gamma\left(\mu_{1}+n\right)\Gamma\left(\mu_{2}+n\right)}{\Gamma(\mu_{1})\Gamma(\mu_{2}) \left(a_{1}a_{2}\right)^{n}}
			{}_{2}F_{1}\left(m_{1},\mu_{1}+n;\mu_{1};c_{1}\right) 
			{}_{2}F_{1}\left(m_{2},\mu_{2}+n;\mu_{2};c_{2}\right) \\
			&= \frac{b_{1}b_{2}\left(\mu_{1}\right)_{n}\left(\mu_{2}\right)_{n}}{ \left(a_{1}a_{2}\right)^{n}}	{}_{2}F_{1}\left(m_{1},\mu_{1}+n;\mu_{1};c_{1}\right){}_{2}F_{1}\left(m_{2},\mu_{2}+n;\mu_{2};c_{2}\right) 
		\end{aligned}
	\end{equation}
	Simplified form of $ n $-th order moment for the case of mixed product, \textit{i.e.}, $ X_{1} $ and $ X_{2} $ follows $ \kappa-\mu  $ shadowed and $ \kappa-\mu $ distribution, respectively, is as follows 
	\begin{equation}
		\begin{aligned}
			\mathbb{E}\left[Y^{n}\right] &= \frac{e^{-\kappa_{2}\mu_{2}} b_{1}\left(\mu_{1}\right)_{n}\left(\mu_{2}\right)_{n}}{ \left(a_{1}a_{2}\right)^{n}}	{}_{2}F_{1}\left(m_{1},\mu_{1}+n;\mu_{1};c_{1}\right){}_{1}F_{1}\left(\mu_{2}+n;\mu_{2};\kappa_{2}\mu_{2}\right) 
			\\ &\stackrel{\left(a\right)}{=} \frac{ b_{1}\left(\mu_{1}\right)_{n}\left(\mu_{2}\right)_{n}}{ \left(a_{1}a_{2}\right)^{n}}	{}_{2}F_{1}\left(m_{1},\mu_{1}+n;\mu_{1};c_{1}\right){}_{1}F_{1}\left(-n;\mu_{2};\kappa_{2}\mu_{2}\right), 
		\end{aligned}
	\end{equation}
	where $ \left(a\right) $ follows from the functional relation given in \cite[Eq. 9.212.1]{Grad2007}. In the next section, we give two applications where these expression are useful.
	\section{Application Examples}\label{Sec:ProductKMS_CommMetrics}
	In a multiple scattering or ``keyholes'' scenario, the wireless channel is modeled as the product of multiple fading distribution \cite{andersen2002statistical,erceg1997comparisons,salo2006statistical,chizhik2002keyholes}. For the case of a double scattered wireless channel, the following are the two application scenario where we considered double $ \kappa-\mu $ shadowed fading channel.
	\subsection{Cascaded Wireless System}
	Consider a two-tap cascaded channel as described in \cite{yilmaz2009product} where both taps follow $ \kappa-\mu $ shadowed fading. This section derives the analytical expression for various important metrics for such a system. 
	\subsubsection{Amount of Fading}
	The amount of fading (AF) measures the severity of any fading channel. It is defined as the ratio of variance to square of the mean of instantaneous SNR \cite{simon2005digital}. Hence, the AF for product $ \kappa-\mu $ shadowed channel is 
	\begin{equation}\label{Eq:AF_Def}
		\begin{aligned}
			\text{AF} &= \frac{\mathbb{V}\left[Y\right]}{\left(\mathbb{E}\left[Y\right]\right)^{2}} = \frac{ \mathbb{E}\left[Y^{2}\right]-\left(\mathbb{E}\left[Y\right]\right)^{2}}{\left(\mathbb{E}\left[Y\right]\right)^{2}}.
		\end{aligned}
	\end{equation}
	By substituting values from (\ref{Eq:Moments}), we have
	\begin{equation}\label{Eq:AFFinal}
		\begin{aligned}
			\text{AF} &=   \left(1 +\frac{2\kappa_{1} + 1}{\mu_{1}\left(1 + \kappa_{1}\right)^{2}} + \frac{\mu_{1}\kappa_{1}^{2} }{m_{1}\left(\mu_{1}+1\right)} \right) \left(1 +\frac{2\kappa_{2} + 1}{\mu_{2}\left(1 + \kappa_{2}\right)^{2}} + \frac{\mu_{2}\kappa_{2}^{2} }{m_{2}\left(\mu_{2}+1\right)} \right) -1. 
		\end{aligned}
	\end{equation}
	The details of the derivation of $ AF $ are presented in Appendix \ref{App:AF}.  
	\subsubsection{Channel Quality Estimation Index}
	In \cite{lioumpas2006channel}, a new performance metric is defined for any wireless channel named as channel quality estimation index (CQEI). By definition, it is the ratio of the variance of the instantaneous SNR to the cube of the mean of instantaneous SNR, \textit{i.e.},
	\begin{equation}\label{Eq:CQEIDef}
		\begin{aligned}
			\text{CQEI}	&= \frac{\mathbb{V}\left[Y\right]}{\left(\mathbb{E}\left[Y\right]\right)^{3}} = \frac{\text{AF}}{\mathbb{E}\left[Y\right]}.
		\end{aligned}
	\end{equation}
	By substituting the value of AF from (\ref{Eq:AFFinal}) in (\ref{Eq:CQEIDef}), we have 
	\begin{equation}
		\begin{aligned}
			\text{CQEI}	&= \frac{1}{\bar{\gamma}_{1}\bar{\gamma}_{2}}\left[\left(1 +\frac{2\kappa_{1} + 1}{\mu_{1}\left(1 + \kappa_{1}\right)^{2}} + \frac{\mu_{1}\kappa_{1}^{2} }{m_{1}\left(\mu_{1}+1\right)} \right) \left(1 +\frac{2\kappa_{2} + 1}{\mu_{2}\left(1 + \kappa_{2}\right)^{2}} + \frac{\mu_{2}\kappa_{2}^{2} }{m_{2}\left(\mu_{2}+1\right)} \right) -1\right]
		\end{aligned}
	\end{equation}
	It can be observed from the expression of AF and CQEI that both of these metrics are monotonically decreasing with $ \mu_{1},m_{1},\mu_{2} $ and $ m_{2} $, \textit{i.e.,} the severity of the channel decreases as these parameter increases. This fact can also be mathematically confirmed by taking the derivative of AF and CQEI. Also, the metrics are are monotonically decreasing with $ \kappa_{1} $ and $ \kappa_{2} $ when $ m_{1} > \mu_{1} $ but monotonically increasing otherwise.
	\subsubsection{Outage Probability}
	In any communication syatem, outage is an event when the received strength of signal falls below a certain threshold. The outage probability (OP) is defined as $ P_{OP}\left(\gamma_{th}\right) = \mathbb{P}\left(Y \le \gamma_{th}\right) $, hence, from (\ref{Eq:ExactCDFMellin1}) we have
	\begin{equation}
		\begin{aligned}
			P_{OP}\left(\gamma_{th}\right)	&= \frac{b_{1}b_{2}}{\Gamma(\mu_{1})\Gamma(\mu_{2})} \sum_{n=0}^{\infty}\left[ \frac{K_{1,n}}{n+\mu_{1}}\gamma_{th}^{n+\mu_{1}} +\frac{K_{2,n}}{n+\mu_{2}}\gamma_{th}^{n+\mu_{2}}  
			\right].
		\end{aligned}
	\end{equation}
	\subsection{Relay with Variable Gain}
	Consider a relay assisted wireless system with a source node ($\mathbf{S}$) communicating with a destination node ($\mathbf{D}$) using a relay ($ \mathbf{R} $). The direct link between source and receiver is assumed be in permanent outage. Hence, the source passes the information signal to relay which amplify-and-forward (AF) it to destination. Here, relay, $ \mathbf{S} $ and $ \mathbf{D} $ are equipped with a single antenna. Assume that $ \mathbf{S} - \mathbf{R} $ link experiences $ \kappa-\mu $ shadowed fading and the $ \mathbf{R} - \mathbf{D} $ link experiences  a cascaded $ \kappa - \mu  $ shadowed fading. Signal-to-noise-ratio (SNR) at $ \mathbf{D} $ with AF-based relay and variable gain is given as \cite{sonia2012VGrelay,zedini2014performance} 
	\begin{equation}
		\begin{aligned}
			\gamma	&=  \frac{\gamma_{sr} \gamma_{rd}}{\gamma_{sr} + \gamma_{rd} + 1} \approx \min\left(\gamma_{sr},\gamma_{rd}\right),
		\end{aligned}
	\end{equation}
	where$ \gamma_{sr},\gamma_{rd} $ represent the SNR of $ \mathbf{S} - \mathbf{R} $ and $ \mathbf{R} - \mathbf{D} $ link respectively. The OP at
	node $ \mathbf{D} $ is evaluated as,
	\begin{equation}\label{Eq:VGR_Outage}
		\begin{aligned}
			P_{OP}^{VGR}\left(\gamma_{th}\right) &= \mathbb{P}\left(\min\left(\gamma_{sr},\gamma_{rd}\right) \le \gamma_{th}\right) \\
			&= F_{\gamma_{sr}}\left(\gamma_{th}\right) +  F_{\gamma_{rd}}\left(\gamma_{th}\right) -  F_{\gamma_{sr}}\left(\gamma_{th}\right) F_{\gamma_{rd}}\left(\gamma_{th}\right),
		\end{aligned}
	\end{equation} 
	where $ F_{\gamma_{sr}}\left(\cdot\right) ,  F_{\gamma_{rd}}\left(\cdot\right) $ are given by (\ref{Eq:ExactCDFMellin1}) or (\ref{Eq:ExactCDFMellin2}) depending on the values of parameters.
	\section{Numerical Results}\label{Sec:Sec:ProductKMS_Results}
\begin{figure}[h]
		\begin{minipage}[b]{0.45\linewidth}
			\centering
			\includegraphics[width=\textwidth]{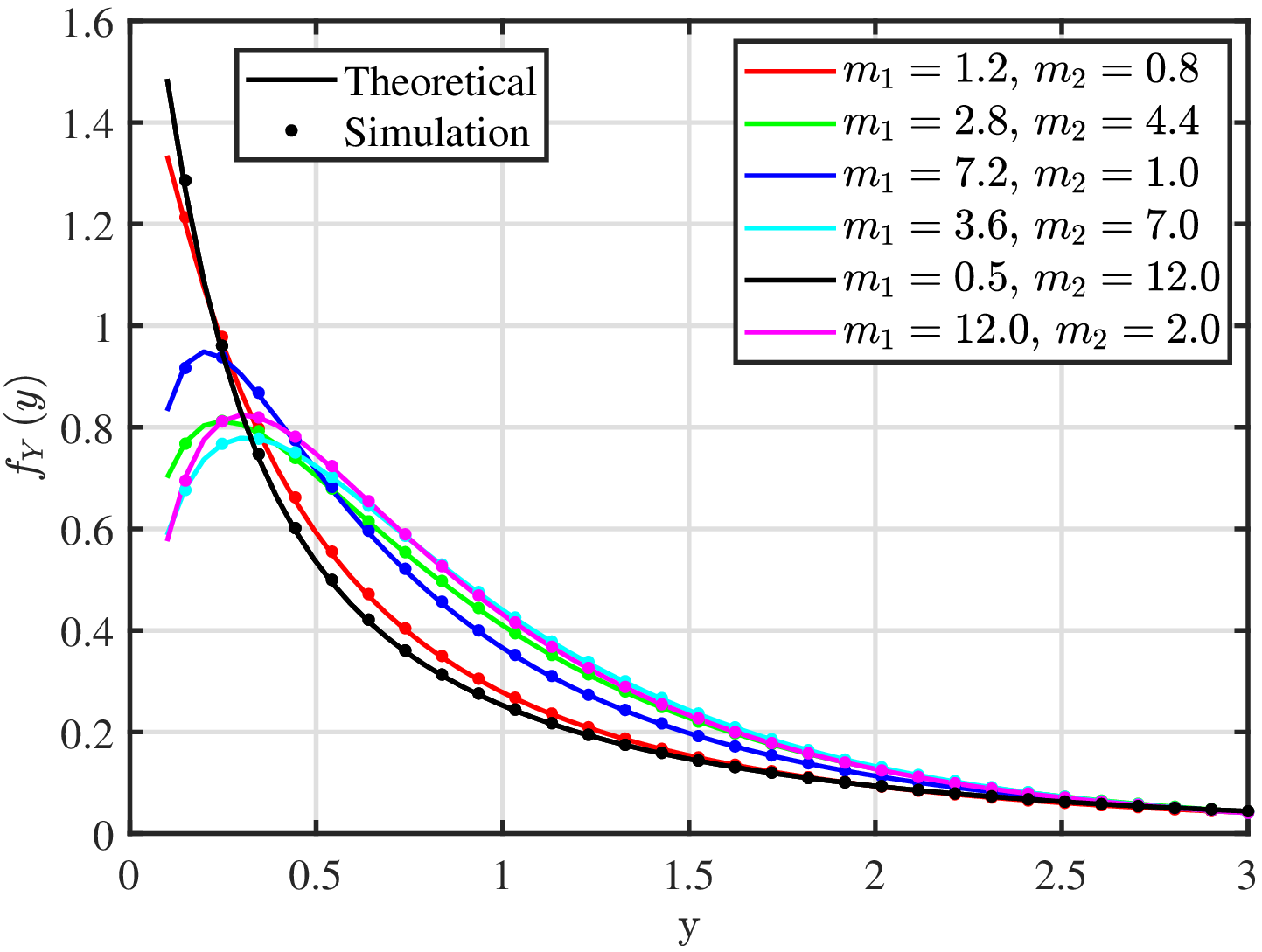}
			\caption{PDF of $ Y $ with $ \{\kappa_{1},\mu_{1} \} = \{5.0,1.2\}$ , $ \{\kappa_{2},\mu_{2}\} = \{2.1,3.0\} $ and various values of $ m_{1} $ and $ m_{2} $}
			\label{Fig:PDF_KMS_Sim_Theo_various_m}
		\end{minipage}
		\hspace{0.5cm}
		\begin{minipage}[b]{0.45\linewidth}
			\centering
			\includegraphics[width=\textwidth]{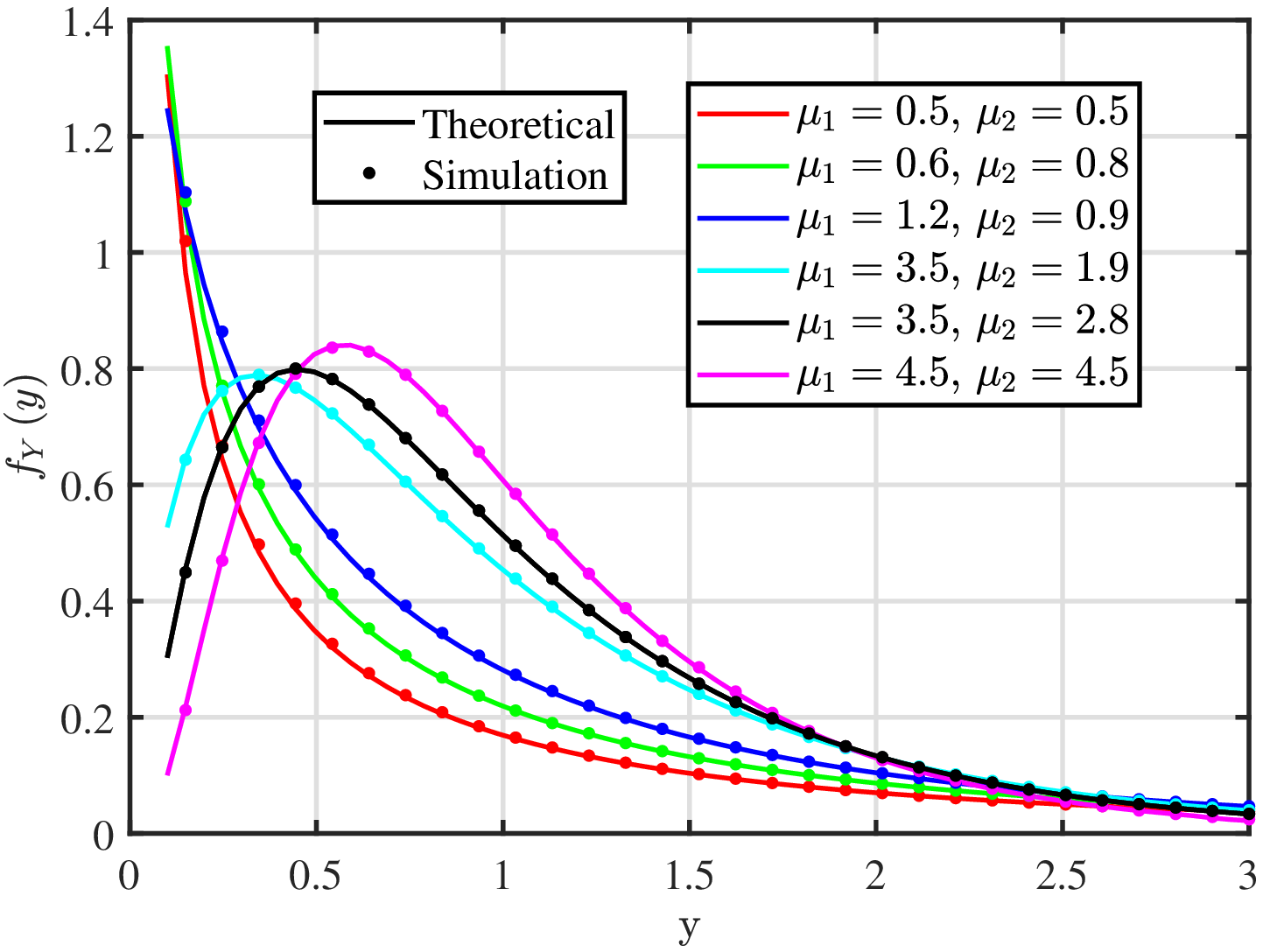}
			\caption{PDF of $ Y $ with $ \{\kappa_{1},m_{1}\} = \{2.2,10\}$ , $ \{\kappa_{2},m_{2}\} = \{0.9,4.0\} $ and various values of $ \mu_{1} $ and $ \mu_{2} $}
			\label{Fig:PDF_KMS_Sim_Theo_various_mu}
		\end{minipage}
	\end{figure}

\begin{figure}[h]
		\begin{minipage}[b]{0.45\linewidth}
			\centering
			\includegraphics[width=\textwidth]{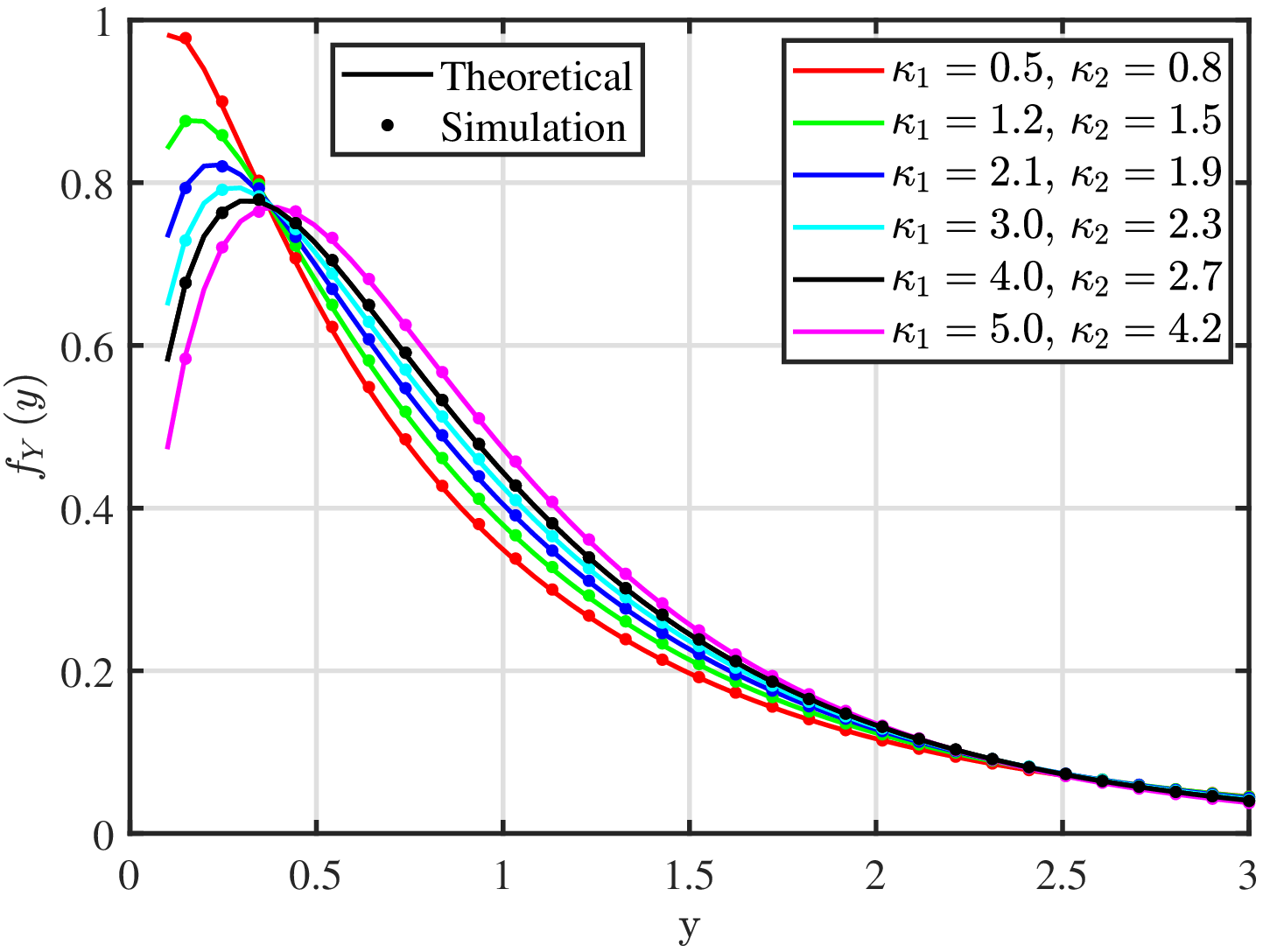}
			\caption{PDF of $ Y $ with $ \{\mu_{1},m_{1}\} = \{2.1,10\}$ , $ \{\mu_{2},m_{2}\} = \{1.5,4.0\} $ and various values of $ \kappa_{1} $ and $ \kappa_{2} $}
			\label{Fig:PDF_KMS_Sim_Theo_various_kappa}
		\end{minipage}
		\hspace{0.5cm}
		\begin{minipage}[b]{0.45\linewidth}
			\centering
			\includegraphics[width=\textwidth]{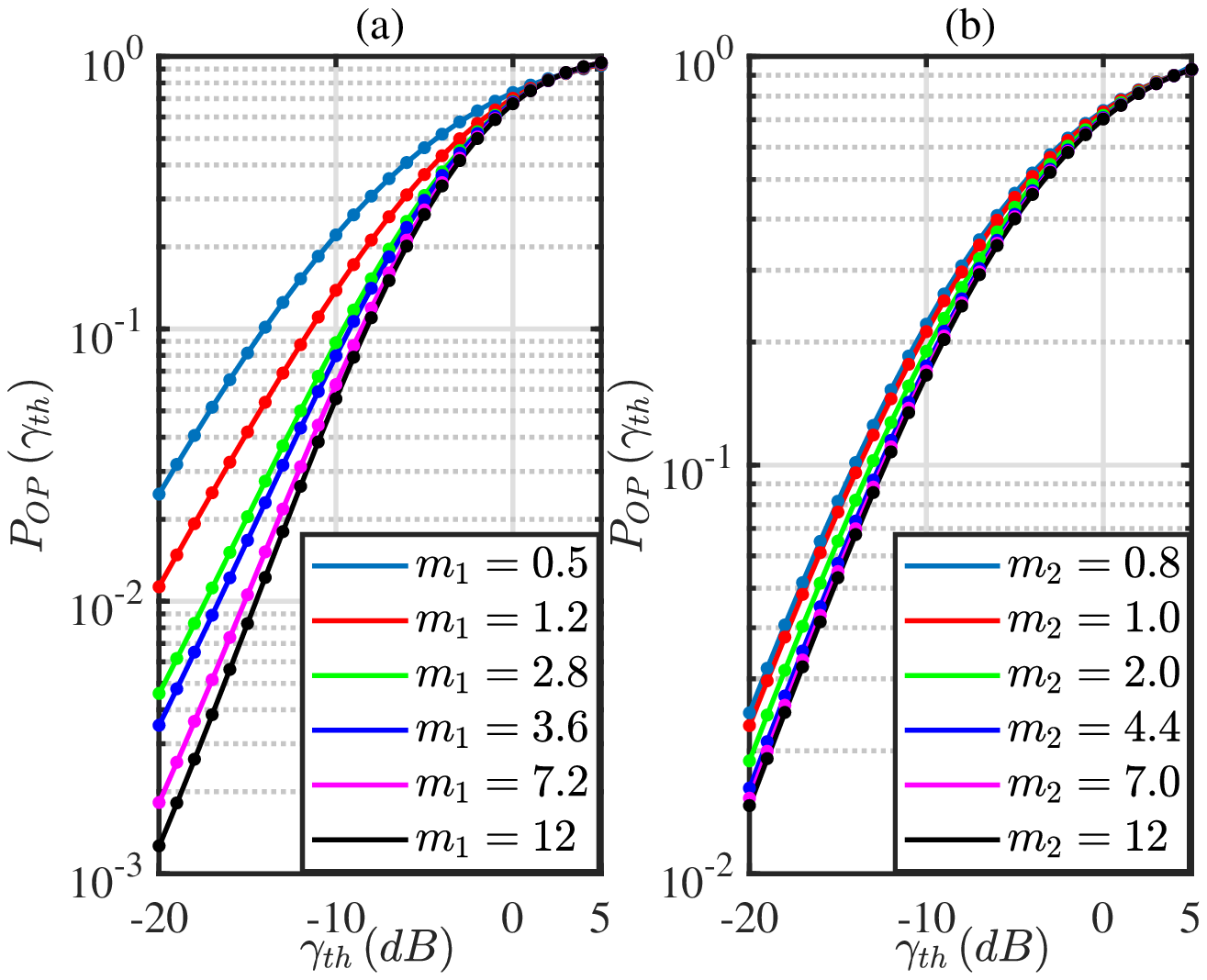}
			\caption{OP of cascaded $ \kappa-\mu $ shadowed channel with $ \{\kappa_{1},\mu_{1} \} = \{5.0,1.2\}$ , $ \{\kappa_{2},\mu_{2} \} = \{2.1,3.0\} $ and various values of $ m_{1} $ and $ m_{2} $}
			\label{Fig:OP_ProdKMS_Sim_Theo_various_m}
		\end{minipage}
	\end{figure}

	This section presents the simulation results that show the correctness and utility of the theoretical expression presented in the previous sections.  Without loss of generality, we have assumed $ \bar{\gamma}_{1} = \bar{\gamma}_{2} = 1 $ for all the plots. In all the figures, We have used solid lines to draw the theoretical values and the dotted markers are used for simulated values. In Figs. \ref{Fig:PDF_KMS_Sim_Theo_various_m}-\ref{Fig:PDF_KMS_Sim_Theo_various_kappa}, several plots for the product PDF of two $ \kappa-\mu $ shadowed RV for various values of $ \kappa,\mu $ and $ m $.  A wide range of shapes can be obtained via choosing different parameter as confirmed by the Figs. \ref{Fig:PDF_KMS_Sim_Theo_various_m}-\ref{Fig:PDF_KMS_Sim_Theo_various_kappa}. Also, one can observe that the simulated PDFs are perfectly matching with the values obtained through theoretical expressions in (\ref{Eq:ExactPDFMellin1}) or (\ref{Eq:ExactPDFMellin2}) when the difference of $ \mu_{2} $ and $ \mu_{1} $ is an integer. It validates the exactness of series expression. 
	
\begin{figure}[h]
		\begin{minipage}[b]{0.45\linewidth}
			\centering
			\includegraphics[width=\textwidth]{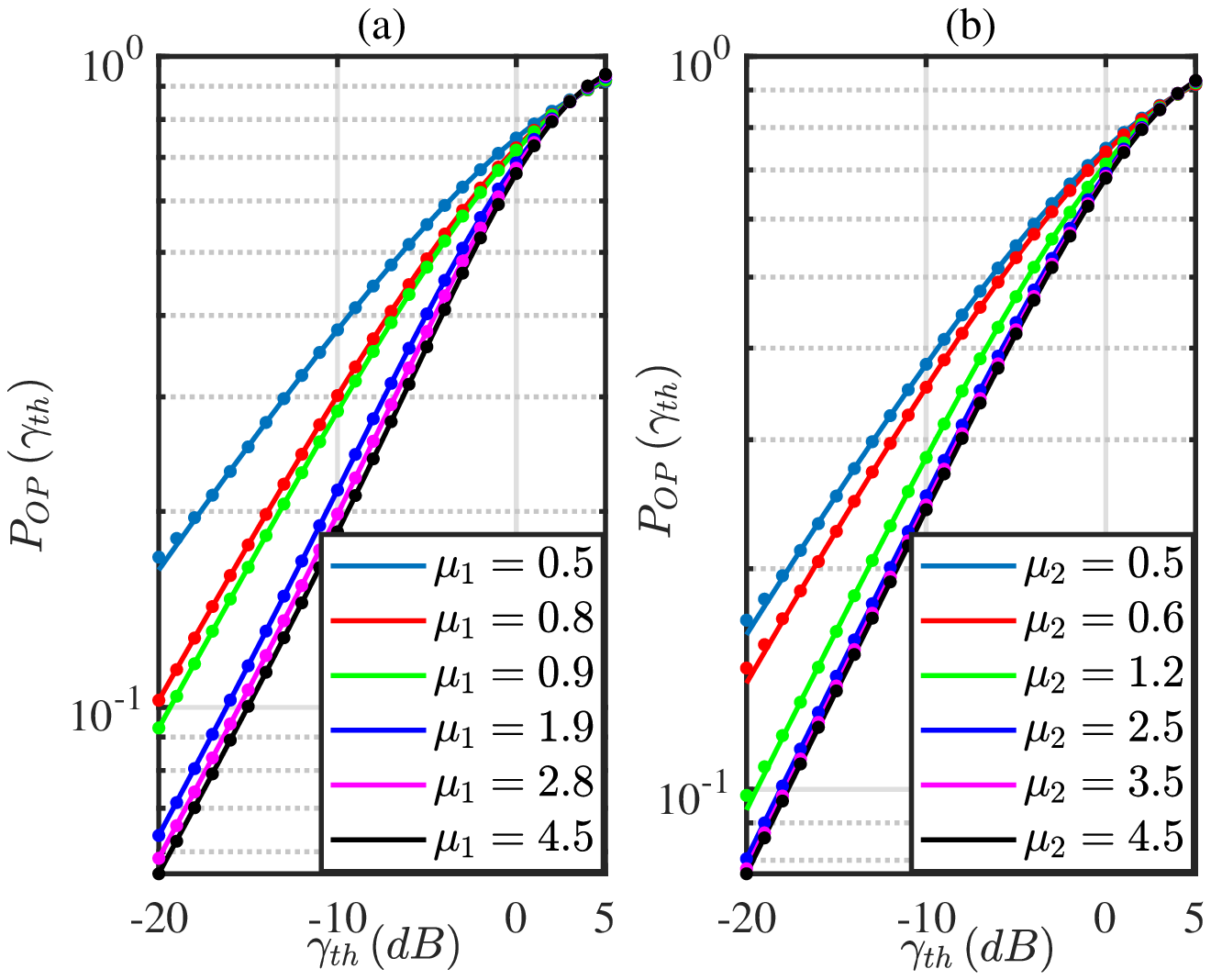}
			\caption{OP of cascaded $ \kappa-\mu $ shadowed channel with  $ \{\kappa_{1},m_{1} \} = \{0.9,4\}$ , $ \{\kappa_{2},m_{2}\} = \{2.2,10\} $ and various values of $ \mu_{1} $ and $ \mu_{2} $}
			\label{Fig:OP_ProdKMS_Sim_Theo_various_mu}
		\end{minipage}
		\hspace{0.5cm}
		\begin{minipage}[b]{0.45\linewidth}
			\centering
			\includegraphics[width=\textwidth]{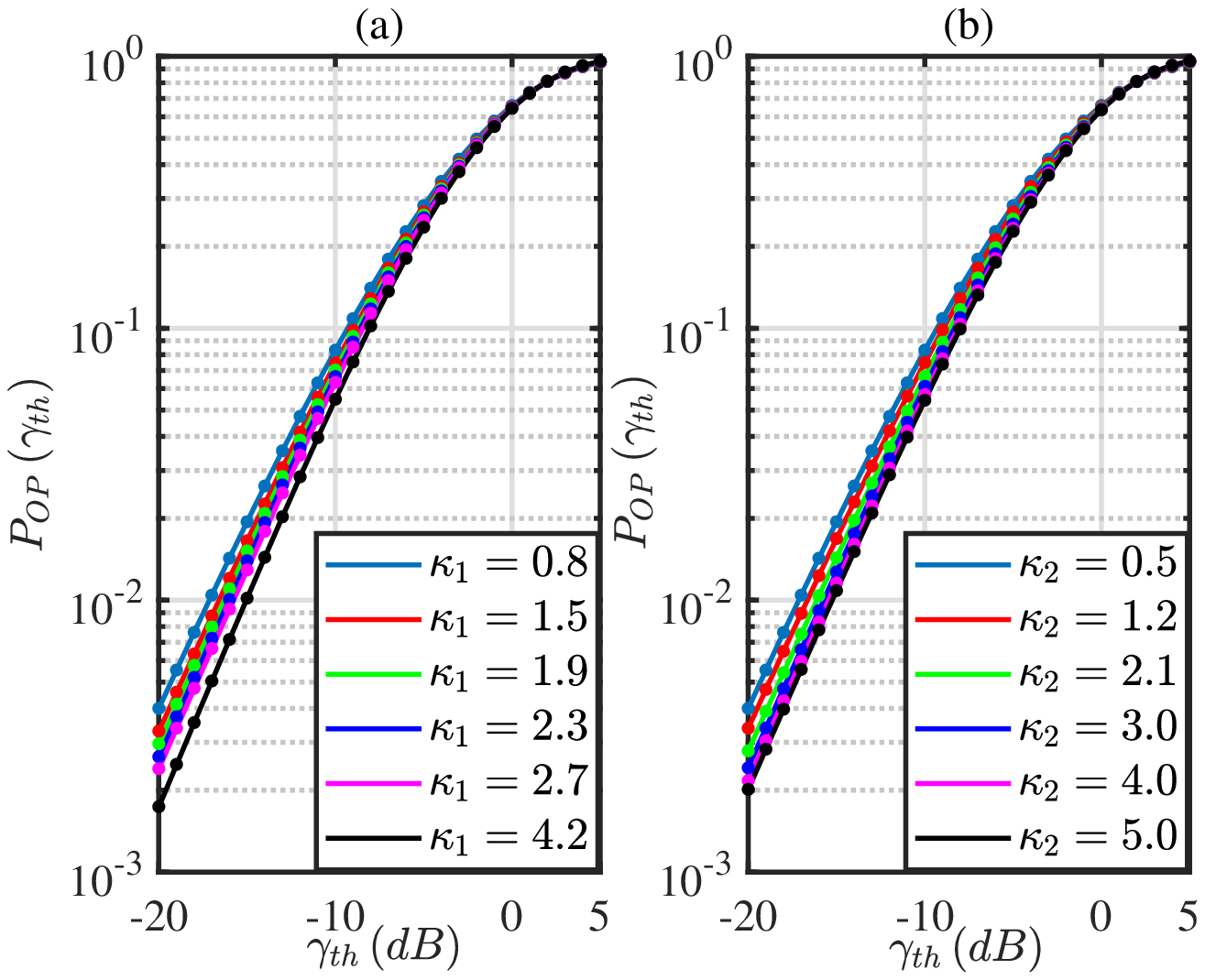}
			\caption{OP of cascaded $ \kappa-\mu $ shadowed channel with $ \{\mu_{1},m_{1} \} = \{1.5,4\}$ , $ \{\mu_{2},m_{2} \} = \{2.1,10\} $ and various values of $ \kappa_{1} $ and $ \kappa_{2} $}
			\label{Fig:OP_ProdKMS_Sim_Theo_various_kappa}
		\end{minipage}
	\end{figure}

	Next, we studied the effect of individual parameters on the OP of a cascaded $ \kappa-\mu $ shadowed fading channel. Fig. \ref{Fig:OP_ProdKMS_Sim_Theo_various_m} shows the impact of increasing the $ m $ parameter of single link when all other parameters are kept constant. In Fig \ref{Fig:OP_ProdKMS_Sim_Theo_various_m}(a), we kept $ \kappa_{1} = 5.0, \mu_{1} = 1.2 $ and $ \kappa_{2} = 2.1, \mu_{2} = 3.0, m_{2} = 0.8 $ then we changed the value of $ m_{1} $.  Similarly in Fig. \ref{Fig:OP_ProdKMS_Sim_Theo_various_m}(b) we kept $ \kappa_{1} = 5.0, \mu_{1} = 1.2, m_{1} = 0.5 $ and $ \kappa_{2} = 2.1, \mu_{2} = 3.0 $ then we changed the value of $ m_{2} $. It can be observed that as $ m $ increases the OP decreases but the effect is not independent of other parameter as in Fig. \ref{Fig:OP_ProdKMS_Sim_Theo_various_m}(a) the impact of increasing $ m $ is more dominant compare to Fig. \ref{Fig:OP_ProdKMS_Sim_Theo_various_m}(b) where it saturates for $ m_{2} = 4.4 $ only. One reason for this behavior may be that the $ \kappa_{1} > \kappa_{2}$ and dominates the overall link. We fixed all the parameter except $ \mu_{1} $  and $ \mu_{2} $ in \ref{Fig:OP_ProdKMS_Sim_Theo_various_mu}(a) and \ref{Fig:OP_ProdKMS_Sim_Theo_various_mu}(b), respectively. Again, we can conclude that as $ \mu $ increases the OP decreases. In other words, a fading channel with higher $ \mu $ is more reliable. In the same way, the impact of $ \kappa $ has been demonstrated in Fig. \ref{Fig:OP_ProdKMS_Sim_Theo_various_kappa}.
	
	\begin{figure}[h]
		\begin{minipage}[b]{0.45\linewidth}
			\centering
			\includegraphics[width=\textwidth]{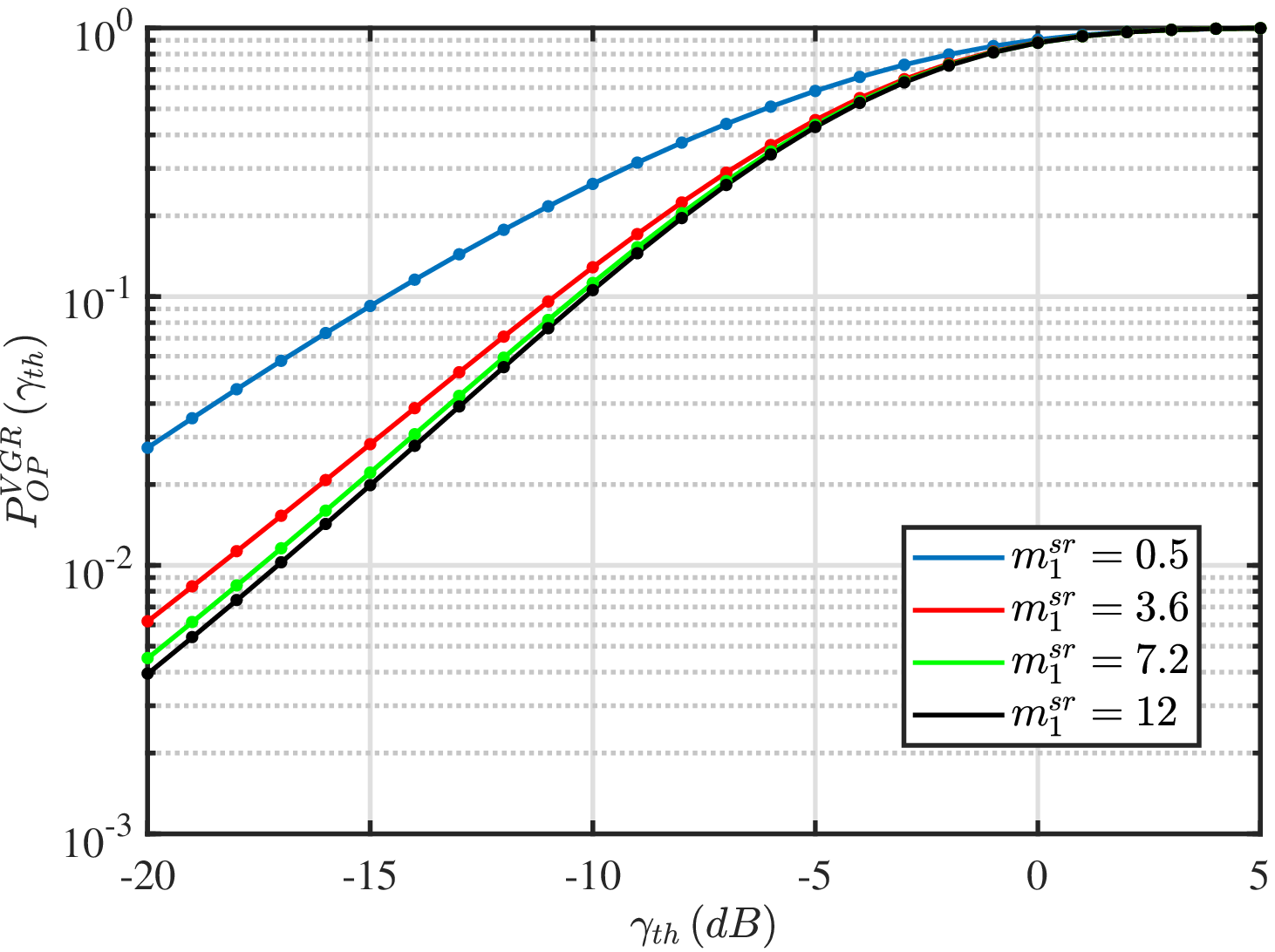}
			\caption{OP for variable relay gain with $ \kappa-\mu $ shadowed fading for various $ m_{1}^{sr} $}
			\label{Fig:VGROP_KMS_Sim_Theo_various_m1}
		\end{minipage}
		\hspace{0.5cm}
		\begin{minipage}[b]{0.45\linewidth}
			\centering
			\includegraphics[width=\textwidth]{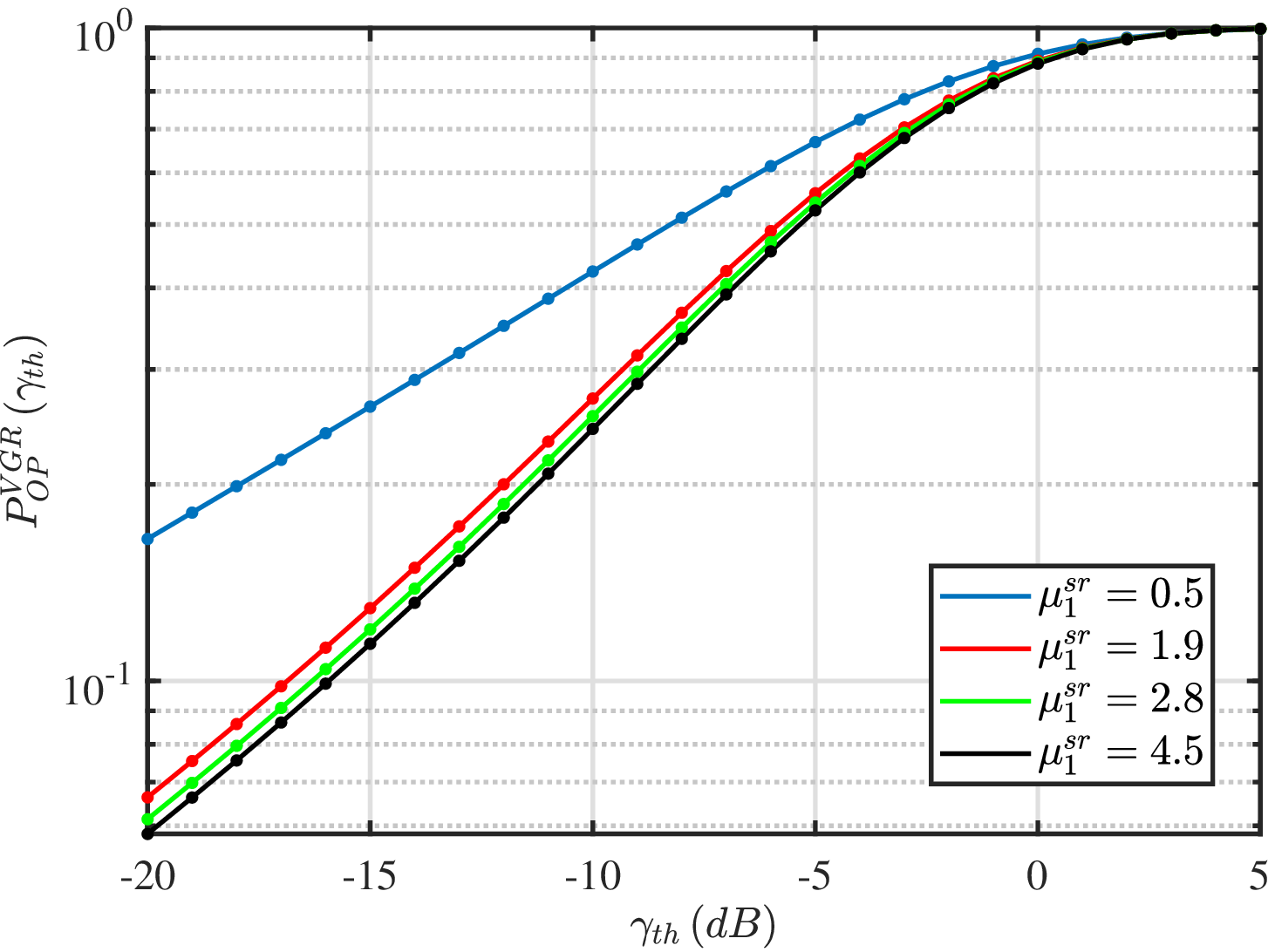}
			\caption{OP for variable relay gain with $ \kappa-\mu $ shadowed fading for various $ \mu_{1}^{sr} $}
			\label{Fig:VGROP_KMS_Sim_Theo_various_mu1}
		\end{minipage}
	\end{figure}	
	
	Next, in Fig. \ref{Fig:VGROP_KMS_Sim_Theo_various_m1} and \ref{Fig:VGROP_KMS_Sim_Theo_various_mu1} we plotted the OP for the relay-assisted wireless communication system with variable gain relay to validate the expression in (\ref{Eq:VGR_Outage}). These figures re-validate the exactness of formulation as the simulated and theoretical values match perfectly. Here, also one can observe that with the increase in parameter values fading channel gets more reliable as the OP decreases.
	
	\section{Conclusion}\label{Sec:ProductKMS_Conclusion}
	This paper presents the series expression for PDF, CDF, and MGF of the product of two $ \kappa-\mu $ shadowed RV. The series expression is obtained via a direct application of Mellin transformation and can be easily computed using popular software like Mathematica. A couple of application examples are also provided for the considered cascaded fading channel.   An interesting extension for this work can be to consider the product of an arbitrary number of $ \kappa-\mu $ shadowed RVs.
	\begin{appendices}
		\section{Evaluation of $ S_{2,n} $}\label{App:ProductKMS_S2}
		\begin{equation}
			\begin{aligned}
				S_{2,n} &= \lim_{s \rightarrow -n-\mu_{1}+1}
				\frac{d}{ds}\Bigg\{
				\frac{\left(s+n+\mu_{1}-1\right)^{2}\Gamma\left(s+\mu_{1}-1\right)\Gamma\left(s+\mu_{2}-1\right)}{ \left(a_{1}a_{2}\right)^{s-1}} 
				{}_{2}F_{1}\left(m_{1},s+\mu_{1}-1;\mu_{1};c_{1}\right) 
				\\&
				\times {}_{2}F_{1}\left(m_{2},s+\mu_{2}-1;\mu_{2};c_{2}\right) y^{-s}
				\Bigg\}  \\
				&= \lim_{s \rightarrow -n-\mu_{1}+1}\curlybracket*{ \phi^{\prime}\left(s\right) h\left(s\right) + \phi\left(s\right) h^{\prime}\left(s\right) } \\
				&= \left(\lim_{s \rightarrow -n-\mu_{1}+1}\phi^{\prime}\left(s\right) \right) \left(\lim_{s \rightarrow -n-\mu_{1}+1} h\left(s\right) \right) + \left(\lim_{s \rightarrow -n-\mu_{1}+1}\phi\left(s\right) \right) \left(\lim_{s \rightarrow -n-\mu_{1}+1}h^{\prime}\left(s\right) \right)
			\end{aligned}
		\end{equation}
		where,
		\begin{equation}
			\begin{aligned}
				\phi\left(s\right)	&=  \left(s+n+\mu_{1}-1\right)^{2}\Gamma\left(s+\mu_{1}-1\right)\Gamma\left(s+\mu_{2}-1\right)
			\end{aligned}
		\end{equation}
		and
		\begin{equation}
			\begin{aligned}
				h\left(s\right)	&= 
				\frac{1}{ \left(a_{1}a_{2}\right)^{s-1}} 
				{}_{2}F_{1}\left(m_{1},s+\mu_{1}-1;\mu_{1};c_{1}\right) 
				{}_{2}F_{1}\left(m_{2},s+\mu_{2}-1;\mu_{2};c_{2}\right) y^{-s}
			\end{aligned}
		\end{equation}
		Following the procedure as described in \cite{mathai1993:handbookGfunction}, we can rewrite $ \phi\left(s\right) $ as 
		\begin{equation}\label{Eq:PhiRewritten}
			\begin{aligned}
				\phi\left(s\right)	
				&= \frac{\Gamma^{2}\left(s+\mu_{1}+n\right) }{\left(s+\mu_{1}+n-2\right)^{2}\dots \left(s+\mu_{1}-1+N\right)^{2} \left(s+\mu_{1}+N-2\right)\dots \left(s+\mu_{1}-1\right)}
			\end{aligned}
		\end{equation}
		From (\ref{Eq:PhiRewritten}), we have
		\begin{equation}\label{Eq:PhiLimit}
			\begin{aligned}
				\lim_{s \rightarrow -n-\mu_{1}+1}\phi\left(s\right)	
				&= \frac{\left(-1\right)^{N}}{\left(n - N\right)! n!}
			\end{aligned}
		\end{equation}
		To compute the $ \phi^{\prime}\left(s\right) $, we take the logarithmic derivative \textit{i.e.,}
		\begin{equation}\label{Eq:DPhi}
			\begin{aligned}
				\phi^{\prime}\left(s\right)	&=  \phi\left(s\right) \frac{d}{ds}\ln\left(\phi\left(s\right)\right)
			\end{aligned}
		\end{equation}
		After some algebraic manipulations, we get
		\begin{equation}
			\begin{aligned}
				\frac{d}{ds}\ln\left(\phi\left(s\right)\right)	&= 2\psi\left(s + \mu_{1} + n\right) - \frac{2}{\left(s+\mu_{1}+n-2\right)} - \dots- \frac{2}{\left(s+\mu_{1}+N-1\right)}
				\\&- \frac{1}{\left(s+\mu_{1}+N-2\right)} - \dots - \frac{1}{\left(s+\mu_{1}-1\right)}.
			\end{aligned}
		\end{equation} 
		\begin{equation}
			\begin{aligned}
				\lim_{s \rightarrow -n-\mu_{1}+1} \frac{d}{ds}\ln\left(\phi\left(s\right)\right)	&= 2 \psi\left(1\right) + 2\left(1 + \frac{1}{2} + \dots + \frac{1}{n - N}\right) + \left(\frac{1}{n - N + 1} + \dots + \frac{1}{n} \right)
			\end{aligned}
		\end{equation} 
		By using \cite[Eq. 1.4.6]{mathai1993:handbookGfunction}, we have
		\begin{equation}\label{Eq:DLogPhiLimit}
			\begin{aligned}
				\lim_{s \rightarrow -n-\mu_{1}+1} \frac{d}{ds}\ln\left(\phi\left(s\right)\right)	&= \psi\left(n+1\right) + \psi\left(n-N+1\right)
			\end{aligned}
		\end{equation}
		After substituting (\ref{Eq:PhiLimit}) and (\ref{Eq:DLogPhiLimit}) in (\ref{Eq:DPhi}), we get
		\begin{equation}
			\begin{aligned}
				\lim_{s \rightarrow -n-\mu_{1}+1} \phi^{\prime}\left(s\right) &= \frac{\left(-1\right)^{N}}{\left(n - N\right)! n!}\left[\psi\left(n+1\right) + \psi\left(n-N+1\right)\right]
			\end{aligned}
		\end{equation}
		As $ h\left(s\right) $ is analytic function so we can simply substitute $ s = -n-\mu_{1}+1 $ to get
		\begin{equation}
			\begin{aligned}
				\lim_{s \rightarrow -n-\mu_{1}+1} h\left(s\right) &= y^{n+\mu_{1}-1}\left(a_{1}a_{2}\right)^{n+\mu_{1}}
				{}_{2}F_{1}\left(m_{1},-n;\mu_{1};c_{1}\right) 
				{}_{2}F_{1}\left(m_{2},-n+N;\mu_{2};c_{2}\right) 
			\end{aligned}
		\end{equation}
		Now, after taking the derivative of $ h\left(s\right) $ and substituting the $ s = -n-\mu_{1}+1 $, we get
		\begin{equation}
			\begin{aligned}
				\lim_{s \rightarrow -n-\mu_{1}+1} h^{\prime}\left(s\right) &=  {}_{2}F_{1}^{\left(0,1,0,0\right)}\left(m_{1},-n;\mu_{1};c_{1}\right)  {}_{2}F_{1}\left(m_{2},-n+N;\mu_{2};c_{2}\right) \left(a_{1}a_{2}\right)^{n+\mu_{1}} y^{n+\mu_{1}-1} \\
				&+ {}_{2}F_{1}^{\left(0,1,0,0\right)}\left(m_{2},-n+N;\mu_{2};c_{2}\right)    {}_{2}F_{1}\left(m_{1},-n;\mu_{1};c_{1}\right)\left(a_{1}a_{2}\right)^{n+\mu_{1}}  y^{n+\mu_{1}-1} \\
				&- \ln\left(a_{1}a_{2}\right){}_{2}F_{1}\left(m_{1},-n;\mu_{1};c_{1}\right)  {}_{2}F_{1}\left(m_{2},-n+N;\mu_{2};c_{2}\right) \left(a_{1}a_{2}\right)^{n+\mu_{1}}  y^{n+\mu_{1}-1} \\
				&- \ln\left(y\right){}_{2}F_{1}\left(m_{1},-n;\mu_{1};c_{1}\right)  {}_{2}F_{1}\left(m_{2},-n+N;\mu_{2};c_{2}\right) \left(a_{1}a_{2}\right)^{n+\mu_{1}}  y^{n+\mu_{1}-1}
			\end{aligned}
		\end{equation} 
		Finally,
		\begin{equation}
			\begin{aligned}
				S_{2,n} &= \frac{\left(-1\right)^{N} \left(a_{1}a_{2}\right)^{n+\mu_{1}} y^{n+\mu_{1}-1} }{\left(n - N\right)! n!} \Bigg\{ {}_{2}F_{1}^{\left(0,1,0,0\right)}\left(m_{1},-n;\mu_{1};c_{1}\right)  {}_{2}F_{1}\left(m_{2},-n+N;\mu_{2};c_{2}\right) \\
				&+ {}_{2}F_{1}\left(m_{1},-n;\mu_{1};c_{1}\right)  {}_{2}F_{1}^{\left(0,1,0,0\right)}\left(m_{2},-n+N;\mu_{2};c_{2}\right) \\
				&+ \left[\psi\left(n+1\right) + \psi\left(n-N+1\right) - \ln\left(y\right) - \ln\left(a_{1}a_{2}\right)\right] 
				{}_{2}F_{1}\left(m_{1},-n;\mu_{1};c_{1}\right)  {}_{2}F_{1}\left(m_{2},-n+N;\mu_{2};c_{2}\right) \Bigg\}.
			\end{aligned}
		\end{equation}
		\section{Calculation of AF}\label{App:AF}
		From the definition of AF in (\ref{Eq:AF_Def}), we have
		\begin{equation}\label{Eq:AF_Def2}
			\begin{aligned}
				AF	&= \frac{\mathbb{E}\left[Y^{2}\right]}{\left(\mathbb{E}\left[Y\right]\right)^{2}} - 1
			\end{aligned}
		\end{equation}
		We have $ \mathbb{E}\left[Y\right] = \bar{\gamma}_{1}\bar{\gamma}_{2} $ and 
		\begin{equation}
			\begin{aligned}
				\mathbb{E}\left[Y^{2}\right] &=  \frac{b_{1}b_{2}\left(\mu_{1}\right)_{2}\left(\mu_{2}\right)_{2}}{ \left(a_{1}a_{2}\right)^{2}}	{}_{2}F_{1}\left(m_{1},\mu_{1}+2;\mu_{1};c_{1}\right){}_{2}F_{1}\left(m_{2},\mu_{2}+2;\mu_{2};c_{2}\right) 
			\end{aligned}
		\end{equation}
		Using the Euler transformation from \cite[Eq. 1.2.2.2]{exton1976multiple}, we have
		\begin{equation}
			\begin{aligned}
				\hspace{-4mm}{}_{2}F_{1}\left(m_{1},\mu_{1}+2;\mu_{1};c_{1}\right)	&= \frac{1}{\left(1-c_{1}\right)^{m_{1}}} {}_{2}F_{1}\left(m_{1},-2;\mu_{1};\frac{c_{1}}{c_{1}-1}\right) \\
				&\hspace{-35mm}= \left(1-c_{1}\right)^{-m_{1}}\left[1 - \frac{2m_{1}}{\mu_{1}}\frac{c_{1}}{c_{1}-1} + \frac{m_{1}\left(m_{1}+1\right)}{\mu_{1}\left(\mu_{1}+1\right)}\left(\frac{c_{1}}{c_{1}-1}\right)^{2}\right] \\
				&\hspace{-35mm}= \frac{1}{b_{1}}\left[1 +2\kappa_{1} + \frac{\mu_{1}\left(m_{1}+1\right)}{m_{1}\left(\mu_{1}+1\right)}\kappa_{1}^{2} \right]
			\end{aligned}
		\end{equation} 
		\begin{equation}
			\begin{aligned}
				\mathbb{E}\left[Y^{2}\right] &=  \frac{\left(\mu_{1}\right)_{2}\left(\mu_{2}\right)_{2}}{ \left(a_{1}a_{2}\right)^{2}}	\left[1 +2\kappa_{1} + \frac{\mu_{1}\left(m_{1}+1\right)}{m_{1}\left(\mu_{1}+1\right)}\kappa_{1}^{2} \right]\left[1 +2\kappa_{2} + \frac{\mu_{2}\left(m_{2}+1\right)}{m_{2}\left(\mu_{2}+1\right)}\kappa_{2}^{2} \right]
			\end{aligned}
		\end{equation}
		\begin{equation}\label{Eq:AF_Part1}
			\begin{aligned}
				\frac{\mathbb{E}\left[Y^{2}\right]}{\left(\mathbb{E}\left[Y\right]\right)^{2}}	&= \frac{\left(\mu_{1} + 1\right)\left(\mu_{2}+1\right)}{\mu_{1}\mu_{2}\left(\kappa_{1}+1\right)^{2}\left(\kappa_{2}+1\right)^{2}}  \left[1 +2\kappa_{1} + \frac{\mu_{1}\left(m_{1}+1\right)}{m_{1}\left(\mu_{1}+1\right)}\kappa_{1}^{2} \right]\left[1 +2\kappa_{2} + \frac{\mu_{2}\left(m_{2}+1\right)}{m_{2}\left(\mu_{2}+1\right)}\kappa_{2}^{2} \right]
			\end{aligned}
		\end{equation}
		Finally, the substitution of (\ref{Eq:AF_Part1}) in (\ref{Eq:AF_Def2}) and some algebraic manipulation results in (\ref{Eq:AFFinal}) and completes the proof.
	\end{appendices}

	\bibliographystyle{IEEEtran}
	\bibliography{RefKappaMuProd}	

\begin{thebibliography}{10}
\providecommand{\url}[1]{#1}
\csname url@samestyle\endcsname
\providecommand{\newblock}{\relax}
\providecommand{\bibinfo}[2]{#2}
\providecommand{\BIBentrySTDinterwordspacing}{\spaceskip=0pt\relax}
\providecommand{\BIBentryALTinterwordstretchfactor}{4}
\providecommand{\BIBentryALTinterwordspacing}{\spaceskip=\fontdimen2\font plus
\BIBentryALTinterwordstretchfactor\fontdimen3\font minus
  \fontdimen4\font\relax}
\providecommand{\BIBforeignlanguage}[2]{{%
\expandafter\ifx\csname l@#1\endcsname\relax
\typeout{** WARNING: IEEEtran.bst: No hyphenation pattern has been}%
\typeout{** loaded for the language `#1'. Using the pattern for}%
\typeout{** the default language instead.}%
\else
\language=\csname l@#1\endcsname
\fi
#2}}
\providecommand{\BIBdecl}{\relax}
\BIBdecl

\bibitem{alouini2002dual}
M.-S. Alouini and M.~K. Simon, ``Dual diversity over correlated log-normal
  fading channels,'' \emph{IEEE Transactions on Communications}, vol.~50,
  no.~12, pp. 1946--1959, 2002.

\bibitem{abdi1999utility}
A.~Abdi and M.~Kaveh, ``On the utility of gamma pdf in modeling shadow fading
  (slow fading),'' in \emph{1999 IEEE 49th Vehicular Technology Conference
  (Cat. No. 99CH36363)}, vol.~3.\hskip 1em plus 0.5em minus 0.4em\relax IEEE,
  1999, pp. 2308--2312.

\bibitem{Yacoub2007:KappaMu}
M.~D. Yacoub, ``The $\kappa$-$\mu$ distribution and the $\eta$-$\mu$
  distribution,'' \emph{IEEE Antennas and Propagation Magazine}, vol.~49,
  no.~1, pp. 68--81, 2007.

\bibitem{yacoub2007alpha}
M.~D. Yacoub, ``The $\alpha - \mu $ distribution: A physical fading model for
  the stacy distribution,'' \emph{IEEE Transactions on Vehicular Technology},
  vol.~56, no.~1, pp. 27--34, 2007.

\bibitem{paris2013:statistical}
J.~F. Paris, ``Statistical characterization of $ \kappa-\mu $ shadowed
  fading,'' \emph{IEEE Transactions on Vehicular Technology}, vol.~63, no.~2,
  pp. 518--526, 2013.

\bibitem{lopez2017kappa}
F.~J. Lopez-Martinez, J.~F. Paris, and J.~M. Romero-Jerez, ``The $\kappa-\mu$
  shadowed fading model with integer fading parameters,'' \emph{IEEE
  Transactions on Vehicular Technology}, vol.~66, no.~9, pp. 7653--7662, 2017.

\bibitem{talha2011channel}
B.~Talha and M.~Paetzold, ``Channel models for mobile-to-mobile cooperative
  communication systems: A state of the art review,'' \emph{IEEE Vehicular
  Technology Magazine}, vol.~6, no.~2, pp. 33--43, 2011.

\bibitem{lis_renzo_2020}
M.~{Di Renzo}, A.~{Zappone}, M.~{Debbah}, M.~S. {Alouini}, C.~{Yuen}, J.~{de
  Rosny}, and S.~{Tretyakov}, ``Smart radio environments empowered by
  reconfigurable intelligent surfaces: How it works, state of research, and the
  road ahead,'' \emph{IEEE Journal on Selected Areas in Communications},
  vol.~38, no.~11, pp. 2450--2525, 2020.

\bibitem{wu2019towards}
Q.~Wu and R.~Zhang, ``Towards smart and reconfigurable environment: Intelligent
  reflecting surface aided wireless network,'' \emph{IEEE Communications
  Magazine}, vol.~58, no.~1, pp. 106--112, 2019.

\bibitem{Zhang5g}
{Zhang, Jiayi and Björnson, Emil and Matthaiou, Michail and Ng, Derrick Wing
  Kwan and Yang, Hong and Love, David J.}, ``{Prospective {M}ultiple {A}ntenna
  {T}echnologies for {B}eyond 5G},'' \emph{{IEEE Journal on Selected Areas in
  Communications}}, vol.~{38}, no.~{8}, pp. {1637--1660}, {2020}.

\bibitem{fasarakis2015coherent}
N.~Fasarakis-Hilliard, P.~N. Alevizos, and A.~Bletsas, ``Coherent detection and
  channel coding for bistatic scatter radio sensor networking,'' \emph{IEEE
  Transactions on Communications}, vol.~63, no.~5, pp. 1798--1810, 2015.

\bibitem{alevizos2017noncoherent}
P.~N. Alevizos, A.~Bletsas, and G.~N. Karystinos, ``Noncoherent short packet
  detection and decoding for scatter radio sensor networking,'' \emph{IEEE
  Transactions on Communications}, vol.~65, no.~5, pp. 2128--2140, 2017.

\bibitem{alevizos2018multistatic}
P.~N. Alevizos, K.~Tountas, and A.~Bletsas, ``Multistatic scatter radio sensor
  networks for extended coverage,'' \emph{IEEE Transactions on Wireless
  Communications}, vol.~17, no.~7, pp. 4522--4535, 2018.

\bibitem{salo2006distribution}
J.~Salo, H.~M. El-Sallabi, and P.~Vainikainen, ``The distribution of the
  product of independent {R}ayleigh random variables,'' \emph{IEEE Transactions
  on Antennas and Propagation}, vol.~54, no.~2, pp. 639--643, 2006.

\bibitem{karagiannidis2007n}
G.~K. Karagiannidis, N.~C. Sagias, and P.~T. Mathiopoulos, ``$ {N} \ast $
  {N}akagami: a novel stochastic model for cascaded fading channels,''
  \emph{IEEE Transactions on Communications}, vol.~55, no.~8, pp. 1453--1458,
  2007.

\bibitem{yilmaz2009product}
F.~Yilmaz and M.-S. Alouini, ``Product of the powers of generalized nakagami-m
  variates and performance of cascaded fading channels,'' in \emph{GLOBECOM
  2009-2009 IEEE Global Telecommunications Conference}.\hskip 1em plus 0.5em
  minus 0.4em\relax IEEE, 2009, pp. 1--8.

\bibitem{Bhargav2018:ProductKappaMu}
N.~Bhargav, C.~R.~N. da~Silva, Y.~J. Chun, E.~J. Leonardo, S.~L. Cotton, and
  M.~D. Yacoub, ``On the product of two $\kappa$-$\mu$ random variables and its
  application to double and composite fading channels,'' \emph{IEEE
  Transactions on Wireless Communications}, vol.~17, no.~4, pp. 2457--2470,
  2018.

\bibitem{Silva:ProductAlphaKappaEtaMu}
C.~R.~N. da~Silva, E.~J. Leonardo, and M.~D. Yacoub, ``Product of two envelopes
  taken from $\alpha $ - $\mu $ , $\kappa $ - $\mu $ , and $\eta $ - $\mu $
  distributions,'' \emph{IEEE Transactions on Communications}, vol.~66, no.~3,
  pp. 1284--1295, 2018.

\bibitem{Charishma2021OutageIRS}
M.~Charishma, A.~Subhash, S.~Shekhar, and S.~Kalyani, ``Outage probability
  expressions for an irs-assisted system with and without source-destination
  link for the case of quantized phase shifts in $\kappa-\mu$ fading,''
  \emph{IEEE Transactions on Communications}, pp. 1--1, 2021.

\bibitem{kumar2015approximate}
S.~Kumar, ``Approximate outage probability and capacity for $\kappa-\mu$
  shadowed fading,'' \emph{IEEE Wireless Communications Letters}, vol.~4,
  no.~3, pp. 301--304, 2015.

\bibitem{kumar2017outage}
S.~Kumar and S.~Kalyani, ``Outage probability and rate for $\kappa-\mu$
  shadowed fading in interference limited scenario,'' \emph{IEEE Transactions
  on Wireless Communications}, vol.~16, no.~12, pp. 8289--8304, 2017.

\bibitem{chandrasekaran2015performance}
G.~Chandrasekaran and S.~Kalyani, ``Performance analysis of cooperative
  spectrum sensing over $\kappa-\mu$ shadowed fading,'' \emph{IEEE Wireless
  Communications Letters}, vol.~4, no.~5, pp. 553--556, 2015.

\bibitem{bhatnagar2014sum}
M.~R. Bhatnagar, ``On the sum of correlated squared $\kappa-\mu$ shadowed
  random variables and its application to performance analysis of {MRC},''
  \emph{IEEE Transactions on Vehicular Technology}, vol.~64, no.~6, pp.
  2678--2684, 2014.

\bibitem{cotton2015second}
S.~L. Cotton, ``Second-order statistics of $\kappa-\mu$ shadowed fading
  channels,'' \emph{IEEE Transactions on Vehicular Technology}, vol.~65,
  no.~10, pp. 8715--8720, 2015.

\bibitem{srinivasan2018secrecy}
M.~Srinivasan and S.~Kalyani, ``Secrecy capacity of $\kappa-\mu$ shadowed
  fading channels,'' \emph{IEEE Communications Letters}, vol.~22, no.~8, pp.
  1728--1731, 2018.

\bibitem{subhash2019asymptotic}
A.~Subhash, M.~Srinivasan, and S.~Kalyani, ``Asymptotic maximum order statistic
  for sir in $\kappa-\mu$ shadowed fading,'' \emph{IEEE Transactions on
  Communications}, vol.~67, no.~9, pp. 6512--6526, 2019.

\bibitem{bilim2022cascaded}
M.~Bilim, ``Cascaded double fading channels: A case of $\kappa-\mu$
  shadowing,'' \emph{Physical Communication}, p. 101649, 2022.

\bibitem{srivastava1985:multiple}
H.~M. Srivastava and P.~W. Karlsson, \emph{Multiple Gaussian hypergeometric
  series}.\hskip 1em plus 0.5em minus 0.4em\relax Ellis Horwood, 1985.

\bibitem{Grad2007}
I.~S. Gradshteyn and I.~M. Ryzhik, \emph{Table of integrals, series, and
  products}.\hskip 1em plus 0.5em minus 0.4em\relax Academic Press, 2007.

\bibitem{springer1979:RValgebra}
M.~D. Springer, \emph{The algebra of random variables}.\hskip 1em plus 0.5em
  minus 0.4em\relax New York: Wiley, 1979.

\bibitem{Polygamma}
\BIBentryALTinterwordspacing
E.~W. Weisstein, \emph{Polygamma Function}, (accessed Feb 12, 2022). [Online].
  Available: \url{https://mathworld.wolfram.com/PolygammaFunction.html}
\BIBentrySTDinterwordspacing

\bibitem{diff2F1wrtb}
\BIBentryALTinterwordspacing
E.~W. Weisstein, \emph{Gauss Hypergeometric Function $ {}_{2} F_{1} $},
  (accessed Feb 12, 2022). [Online]. Available:
  \url{http://functions.wolfram.com/07.23.20.0004.01}
\BIBentrySTDinterwordspacing

\bibitem{andersen2002statistical}
J.~B. Andersen, ``Statistical distributions in mobile communications using
  multiple scattering,'' in \emph{Proc. 27th URSI General Assembly}, 2002, pp.
  1--4.

\bibitem{erceg1997comparisons}
V.~Erceg, S.~J. Fortune, J.~Ling, A.~Rustako, and R.~A. Valenzuela,
  ``Comparisons of a computer-based propagation prediction tool with
  experimental data collected in urban microcellular environments,'' \emph{IEEE
  Journal on Selected Areas in Communications}, vol.~15, no.~4, pp. 677--684,
  1997.

\bibitem{salo2006statistical}
J.~Salo, H.~M. El-Sallabi, and P.~Vainikainen, ``Statistical analysis of the
  multiple scattering radio channel,'' \emph{IEEE Transactions on Antennas and
  Propagation}, vol.~54, no.~11, pp. 3114--3124, 2006.

\bibitem{chizhik2002keyholes}
D.~Chizhik, G.~J. Foschini, M.~J. Gans, and R.~A. Valenzuela, ``Keyholes,
  correlations, and capacities of multielement transmit and receive antennas,''
  \emph{IEEE Transactions on Wireless Communications}, vol.~1, no.~2, pp.
  361--368, 2002.

\bibitem{simon2005digital}
M.~K. Simon and M.-S. Alouini, \emph{Digital communication over fading
  channels}.\hskip 1em plus 0.5em minus 0.4em\relax John Wiley \& Sons, 2005,
  vol.~95.

\bibitem{lioumpas2006channel}
A.~S. Lioumpas, G.~K. Karagiannidis, and A.~C. Iossifides, ``Channel quality
  estimation index ({CQEI}): An improved performance criterion for wireless
  communications systems over fading channels,'' in \emph{12th European
  Wireless Conference 2006-Enabling Technologies for Wireless Multimedia
  Communications}.\hskip 1em plus 0.5em minus 0.4em\relax VDE, 2006, pp. 1--6.

\bibitem{sonia2012VGrelay}
S.~S. Ikki and S.~Aissa, ``Performance evaluation and optimization of dual-hop
  communication over {N}akagami$ -m $ fading channels in the presence of
  co-channel interferences,'' \emph{IEEE Communications Letters}, vol.~16,
  no.~8, pp. 1149--1152, 2012.

\bibitem{zedini2014performance}
E.~Zedini, I.~S. Ansari, and M.-S. Alouini, ``Performance analysis of mixed
  {N}akagami$-m $ and gamma--gamma dual-hop fso transmission systems,''
  \emph{IEEE Photonics Journal}, vol.~7, no.~1, pp. 1--20, 2014.

\bibitem{mathai1993:handbookGfunction}
A.~M. Mathai, \emph{A handbook of generalized special functions for statistical
  and physical sciences}.\hskip 1em plus 0.5em minus 0.4em\relax Oxford
  University Press, USA, 1993.

\bibitem{exton1976multiple}
H.~Exton, \emph{Multiple hypergeometric functions and applications}.\hskip 1em
  plus 0.5em minus 0.4em\relax Ellis Horwood, 1976.

\end{thebibliography}
\end{document}